\documentclass[aps,prb,twocolumn,floatfix]{revtex4}
\usepackage{amsmath}
\usepackage{amssymb}
\usepackage{graphicx}
\usepackage{dcolumn}
\usepackage{bm}
\usepackage{color}

\begin{document} 

\title{Effects of thickness in quantum dots at strong magnetic field}

\author{E. T\"ol\"o}

\author{A. Harju} 

\affiliation{Helsinki Institute of Physics and Department of Applied Physics, Helsinki University of Technology, P.O. Box 4100, FI-02015 HUT, Finland}

\begin{abstract}
  We study the effects of thickness on the ground states of
  two-dimensional quantum dots in high magnetic fields. To be
  specific, we assume the thickness to be small so that only the
  lowest state in the corresponding direction is occupied, but which however
  leads to a modification of the effective interaction between the electrons. We find the
  ground state phase diagram and demonstrate the emergence of new
  phases as the thickness is accounted for. Finally, the wave
  functional form and vortex structure of different phases is
  analyzed.
\end{abstract}

\maketitle

\section{Introduction}
In recent years, there has been an increasing interest towards two-dimensional quantum dots realized in semiconductor heterostructures. Besides applications in quantum information and quantum computing, quantum dots are interesting in their own right as an example of strongly correlated interacting quantum systems, in which the strong influence of the magnetic field entails prominent diverseness. Moreover, the results extrapolated from computationally feasible few electron droplets are frequently used to understand macroscopic quantum phenomena such as the quantum Hall effects.

\begin{figure}[b]
\includegraphics{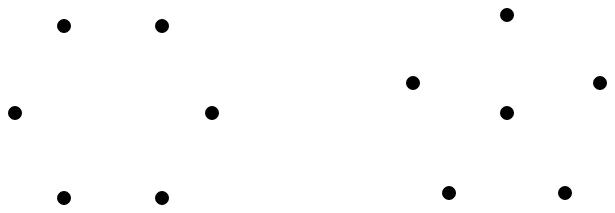}
%
\caption{Likely configurations in six electron quantum dots.}
\label{cfg}
\end{figure}

In this paper, we report an exact diagonalization study of the effects of the layer thickness on the ground states of quantum dots in high magnetic fields. Earlier studies have pointed out \cite{ruan,seki} that in a strictly two-dimensional parabolic quantum dot in  a strong magnetic field and with up to seven electrons, the ground states have strong correlations that favor either of the two classical configurations of the types depicted in Fig.~\ref{cfg},  a regular polygon with or without an electron at the center. This leads to the allowed angular momentum values $M=M_{\rm MDD}+k\Delta M$, where $k$ is a non-negative integer, $\Delta M=N$ or $N-1$ (with $\hbar=1$), and $M_{\rm MDD}=N(N-1)/2$ is the minimum angular momentum of the spin-polarized quantum dot dictated by the Pauli principle. Alternative to the geometric approach is the microscopic composite fermion theory (as opposed to the not so succesful mean-field composite fermion theory), generality of which also extends to larger particle numbers.\cite{jeon}

The vortices in quantum dots with zero thickness have been considered in Ref. \onlinecite{saarikoski_prl}. Taking into account the thickness of the quantum dot is, however, occasionally necessary for understanding of the experimental results.\cite{nishi,nishi2,nazmitdinov} Since the fractional quantum Hall effect is destroyed in thick systems\cite{he},
which relates to unbinding transition of the vortices at the 
electron positions\cite{girvin}, it is of a more general interest to examine, what type of vortex structures are favored at different values of  the thickness parameter.


In Sec.~II, we present the microscopic model and a method to solve it. Sec.~III contains the phase diagrams and related discussion, followed by an analysis of the microscopic states and vortex structures of different phases in Sec.~IV. Finally, Sec.~V summarizes the main results.

\section{Microscopic model}
Quantum dots formed 
in the ${\rm GaAs}/{\rm Al}_{x}{\rm Ga}_{1-x}{\rm As}$
heterostructure are modeled for both lateral and vertical
devices as droplets of electrons in an effectively two-dimensional
plane confined by a harmonic external potential.
We use an effective-mass Hamiltonian 
\begin{equation}
\label{ham}
H=\sum^N_{i=1}\left[
 \frac{({\bf p}_i+\frac{e}{c} {\bf A}_i )^2}{2 m^*}
+\frac{m^*\omega_0^2r_i^2}{2} \right] +  \sum_{i<j}
V(r_{ij})\ ,
\end{equation}
where $N$ is the number of electrons, $m^*=0.067m_e$ is the effective mass and ${\bf A}$ is the vector potential
of the homogeneous  magnetic field ${\bf B}$ perpendicular to the quantum dot plane. 
In the calculations, we set the confinement strength $\hbar\omega_0$ to $2\,{\rm meV}$ as its scaling
should merely shift the ranges of magnetic fields for different phases. For
convenience, we express lengths in units of effective oscillator length $l=\sqrt{\hbar/m^*\omega}$, where
 $\omega=\sqrt{\omega_0^2+(\omega_c/2)^2}$ and  $\omega_c=eB/m^*c$ is the effective cyclotron frequency.

To accommodate the thickness of the electron layer, we employ an effective interaction potential
\begin{equation}
\label{zds}
V(r)=\frac{e^2}{\epsilon}\frac{1}{\sqrt{r^2+d^2}}\ ,
\end{equation}
in which $d$ is comparable to the extent of the wave functions in the $z$-direction and $\epsilon=12.7$ (CGS) is the dielectric constant
of the GaAs semiconductor medium. Typically, an effective interaction is obtained by assuming certain form for the potential in the $z$-direction, solving for its ground state, and integrating out the $z$-direction. While the interaction in Eq. (\ref{zds}) does not relate to any particular wave function, it leads to same qualitative behaviour, especially for smaller $d$, as more realistic effective interactions and is extensively used in the field.\cite{peterson}

\begin{figure*}[t]
\includegraphics{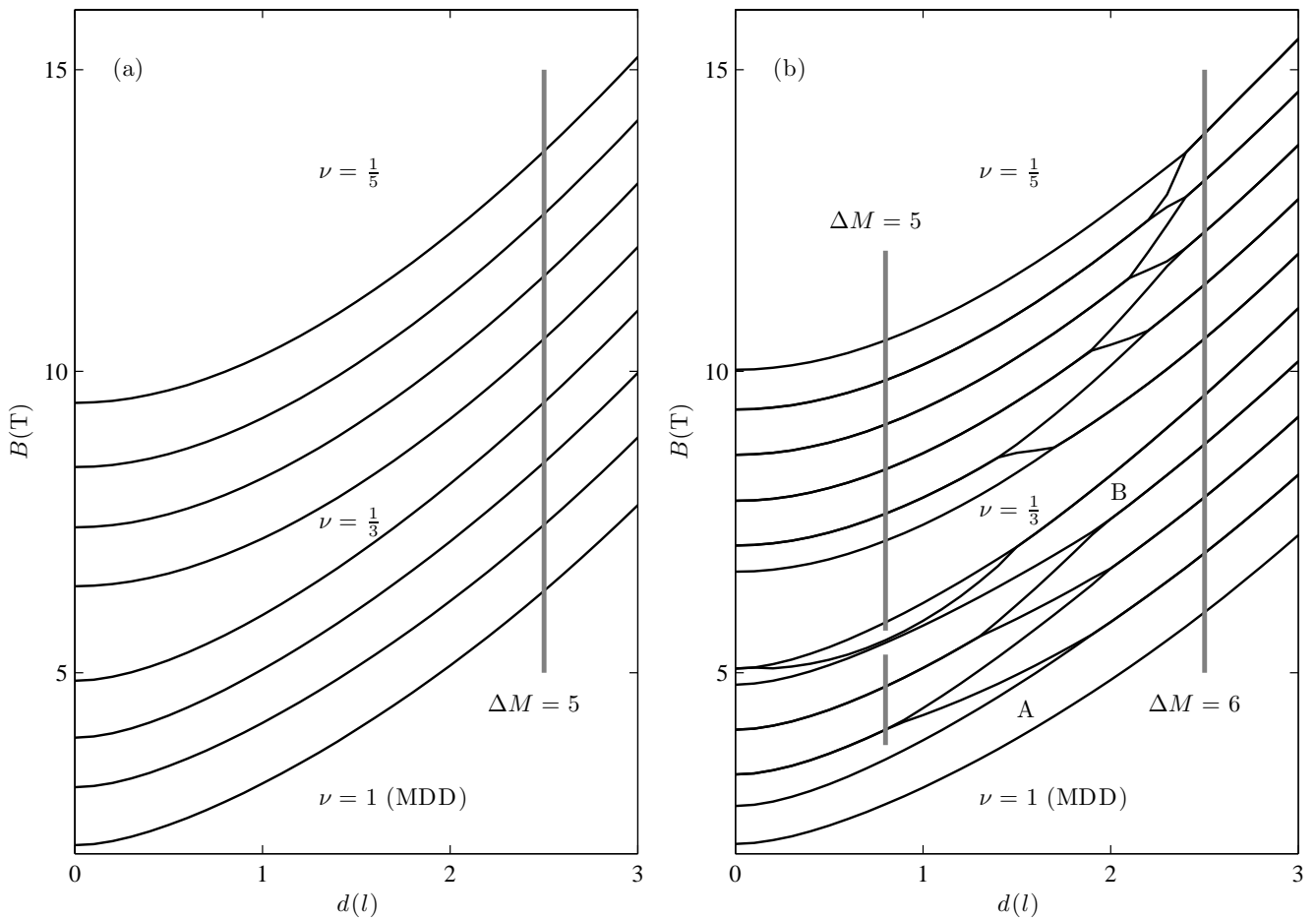}
\caption{The ground state angular momentum phase diagram for (a) five and (b) six electrons as a function of the thickness parameter $d$ and the magnetic field $B$. The vertical lines indicate the regular angular momentum difference $\Delta M$ between adjacent phases. For six electrons, the angular momentum difference changes depending on the thickness except for the phases A and B, as well as for the states with $\nu=1$, $\frac{1}{3}$, and $\frac{1}{5}$.}
\label{fig:56}
\end{figure*}

The ground state of Eq. (\ref{ham}) is solved by constructing the many-body Hamiltonian matrix
in the basis of spin-polarized lowest Landau level (more accurately the lowest Fock-Darwin band since $\omega_0\neq0$) and finding
its lowest eigenstate by the Lanczos diagonalization. The former constitutes a Landau level projection,
an approximation that is valid at the high magnetic field regime.\cite{siljamaki} In the oscillator units, the single-particle wave functions read
\begin{equation}
\label{fd}
\langle z|m\rangle=\tfrac{1}{\sqrt{\pi m!}}z^me^{-z\bar{z}/2}\ , \ \ m\geqslant0\ ,
\end{equation}
where $z=x+iy$. The non-trivial quantities are the interaction matrix elements $\langle m',n'|V(r_{12})|m,n\rangle$. Utilizing the angular momentum conservation $m'+n'=m+n$, these can be written in terms of 
\begin{equation}
M^l_{mn}=\langle m+l,n|V(r_{12})|m,n+l\rangle\ , \ \ l,m,n\geqslant0\ ,
\end{equation}
for which an analytic formula is given in the Appendix.

\section{Phase diagrams}

\begin{figure*}[t]
\includegraphics{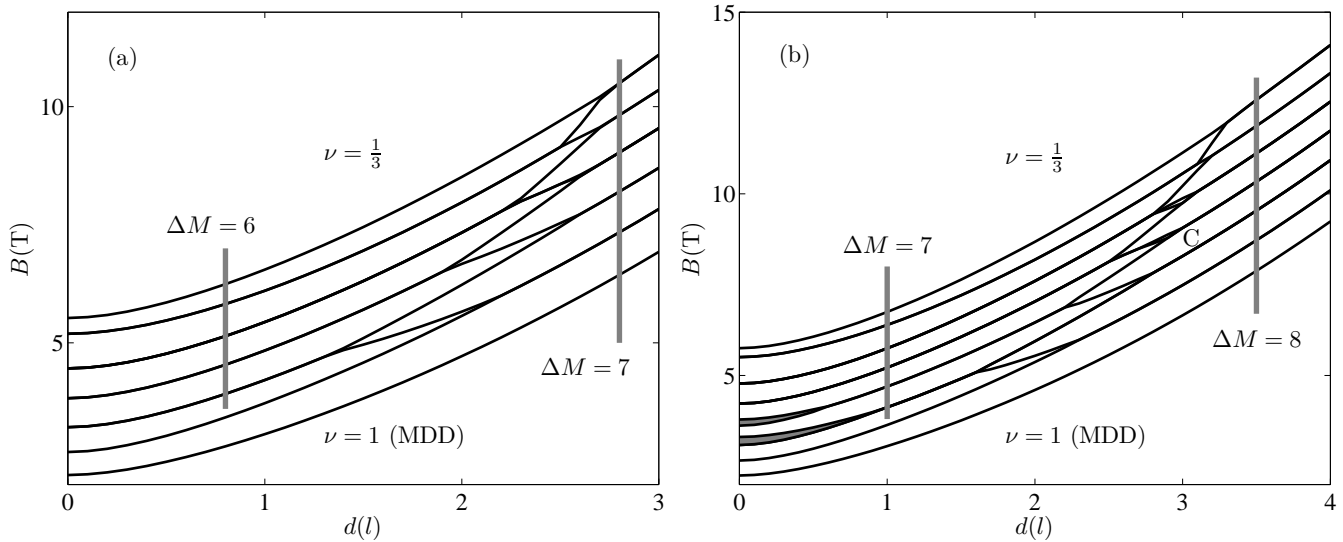}
\caption{Same as Fig.~\ref{fig:56} but for (a) seven and (b) eight electrons.}
\label{fig:78}
\end{figure*}

To obtain a phase diagram, we solve the interaction energy of the ground state for each angular momentum value at each thickness, and determine the angular momentum that has the lowest energy for each pair $(d,B)$ on a grid. Once we have established the angular momenta of neighboring phases, say $M_1$ and $M_2$ for certain $d$, we exactly solve the magnetic field $B$ at the phase boundary by equating the energies $E(d,B,M_1)$ and $E(d,B,M_2)$ and using a standard routine to solve $B$. We limit the considered angular momenta to $M\leqslant M(\nu=\frac{1}{5})$ for five and six electrons and to $M\leqslant M(\nu=\frac{1}{3})$ for seven and eight electrons, where we utilize the stability of the Laughlin states in order to only have stable phases in the phase diagrams. The corresponding angular momenta can be computed from the trial wave functions\cite{laughlin} and are given by
\begin{equation}
\label{la}
M(\nu=\tfrac{1}{k})=\frac{kN(N-1)}{2}\ .
\end{equation}
The single-particle bases are limited according to $m\leqslant k(N-1)$, for the two $k$, which is the basis size needed for the $\nu=\frac{1}{k}$ Laughlin state. Returning back to Fig.~\ref{cfg}, we remark that according to Eq. (\ref{la}) the states that support both classical configurations have the same angular momentum as the Laughlin states.

Fig.~\ref{fig:56} presents the phase diagram that is the ground state angular momentum as a function of the thickness parameter $d$ and the magnetic field $B$ for five and six electrons. If the electron number is less than six, the structure of the phase diagram remains unchanged as the thickness is increased as seen in Fig.~\ref{fig:56}(a) for $N=5$. The adjacent phases are then always separated by angular momentum $\Delta M=N$. In general, the monotonic behaviour of the phase boundaries results from the decrease in the interaction energy as the thickness is increased. In the case of six electrons in Fig.~\ref{fig:56}(b), there is a transition from $\Delta M=N-1$ structure to $\Delta M=N$ structure as the thickness is increased. As an exception, the MDD state with a hole, $M=M_{\rm MDD}+6$ (phase A), has a stable phase all along. Also the phase with $M=M(\nu=\frac{1}{3})-6$ (phase B), being the analogy of the Laughlin's quasielectron wave function, narrowly retains across the whole thickness range as indicated by the break in the vertical line. As a side note, we have established that the overlap between states with the same angular momentum $\langle\psi(d=0)|\psi(d)\rangle$ decreases monotonically but remains significant through the parameter range $d\in[0,3l]$ (the overlap is $\sim0.67$ for $\nu=\frac{1}{3}$ at $d=3l$ and ranges up to $0.98$ for the other states).

The same basic pattern recurs to the systems with seven and eight electrons as seen in Fig.~\ref{fig:78}, where we have confined ourself to the ground states with $M\leqslant M(\nu=\frac{1}{3})$. For eight electrons, there are two additional states with $M=46$ and $M=52$ at small $d$ (shaded in the Fig.~\ref{fig:78}(b)) that appear to be subsequent elements of sequence $M=M_{\rm MDD}+k(N-2)$ with $k=3$ and $4$. Indeed, we have established that the most probable configuration for these states is such that six electrons form a ring around a pair of electrons that resides near the center of the dot. The latter angular momentum, $M=52$, is also equal to that of three holes in the maximum density droplet $M=M_{\rm MDD}+3N$ (phase C found in the diagram after $d\sim2.2l$) and the Pfaffian state\cite{read} $M_{\rm Pf}=\frac{N(2N-3)}{2}$. In Ref. \onlinecite{saarikoski_pf}
, the overlap of the corresponding states with the Pfaffian state was investigated and was found to peak high slightly below $d=2l$ for eight electrons. Interestingly, this ground state angular momentum is unstable when $d\in[0.5l,2.2l]$ indicating that, despite the high overlap, electron pairing is not energetically favorable in the lowest Landau level even if the thickness is optimized. Finally, we should mention the minor phase at $d=3l$ and $B=10$ T with $M=64$ that can be scarcely seen in Fig.~\ref{fig:78}(b) and is of little importance.

With regard to experiments and an estimate for $d$, we have compared the phase diagram for a two-electron quantum dot (figure not shown) to the data presented in Ref. \onlinecite{nazmitdinov} where a full 3D treatment of the system was used to explain experimental results. The magnetic field values at the transition points between the spin-polarized ground states appear to be in a good agreement with the data when the thickness parameter $d\sim10$ nm, which corresponds to about $d\sim l$ at $B=10$ T. Therefore, we expect at least the effects (besides the obvious shifts in $B$) around $d\sim l$ for six and eight electrons to be relevant for an experiment similar to that conducted in Ref. \onlinecite{nishi}, in which the electron number, however, was limited to the maximum of five.

\begin{figure*}[t]
\includegraphics{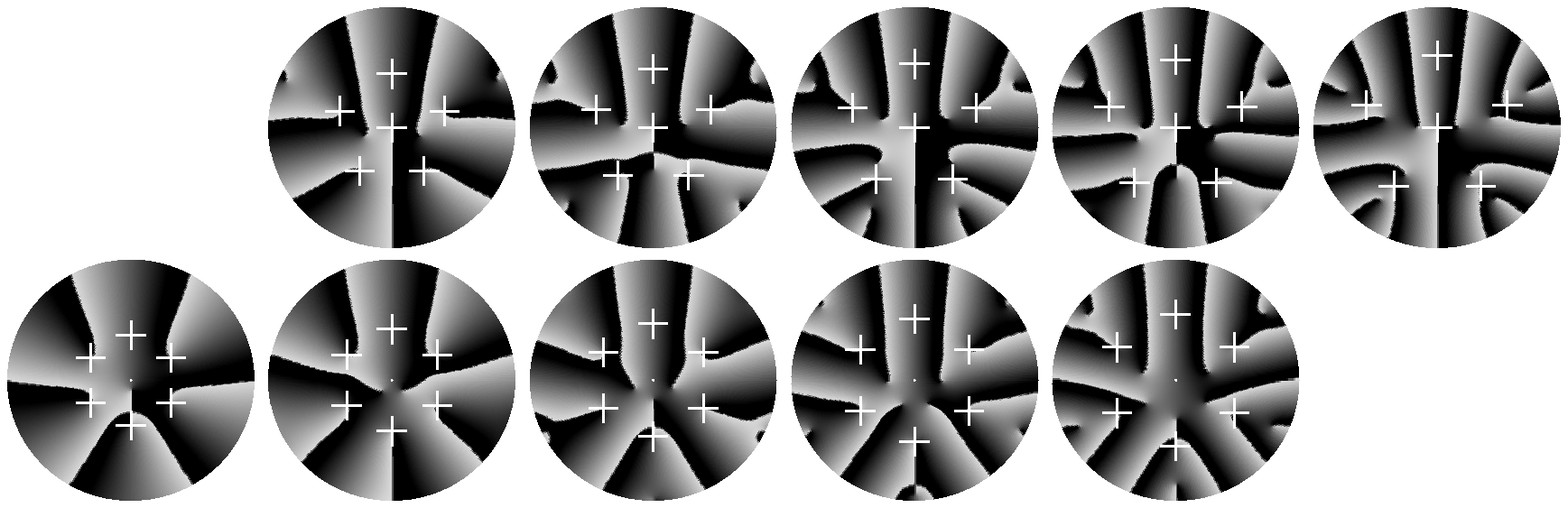}
\caption{The complex phase of the conditional wave functions for the six electron quantum dot, in which the electron co-ordinates are fixed according to the most likely configuration indicated by +, and the upmost electron is used as the probe electron. In the first row, the thickness parameter $d=l$ and the angular momenta are $M=M_{\rm MDD}+5k$ with $k=2,3,4,5,$ and $6$. In the second row, the thickness parameter $d=3l$ and the angular momenta are $M=M_{\rm MDD}+6k$ with $k=1,2,3,4,$ and $5$.}
\label{cwf}
\end{figure*}

At all electron numbers, the range of magnetic field for each phase tends to be equal as the thickness is large. This behaviour also points towards microscopic explanation of these phases by vortex formation as the creation of an additional vortex is associated with a constant flux increase by one flux quantum $\Phi_0=hc/e$.\cite{laughlin}
\section{Vortex structures}

While the values of angular momenta in the phase diagrams are naturally explained by the geometric argument, a more accurate description of the microscopic states can be obtained by investigating the zeros of the wave functions. \cite{stopa} Since the many-body wave function is up to the exponential factor merely an analytic polynomial in the complex electron co-ordinates $z_i$, the zeros are clearly visible in the phase structure of the wave function. For this purpose, we define the phase of the conditional wave function
\begin{equation}
\varphi(z)=\arg \left[\frac{
\Psi(z,z'_2,\ldots,z'_N)
}{
\Psi(z'_1,z'_2,\ldots,z'_N)
}\right]\ ,
\end{equation}
where $z'_i$ denote the electron co-ordinates fixed to chosen positions. Circumvention of each zero, henceforth called a vortex, of $p(z)=\Psi(z,z_2,\ldots,z_N)$ accumulates a factor $2\pi$.


The phases $\varphi$ for the system of six electrons of the ground states up to the angular momentum $M(\nu=\frac{1}{3})$ at $d=l$ and $3l$ are shown in Fig.~\ref{cwf}. The vortex on top of each fixed co-ordinate follows from the necessary presence of the factor $\prod(z_i-z_j)$ in each solution $\Psi$. Of special interest are the free vortices in the central area of the quantum dot, which typically have an irrelevant reflection vortex outside the polygon pattern. In the upper row of the figure, an angular momentum increase by $\Delta M=5$ always leads to an additional vortex near the center, while on the lower row an increase of $\Delta M=6$ is required. Similar results hold for the larger angular momenta and electron numbers.

\begin{table}[t]
\caption{Overlaps with the trial wave functions for $\Delta M=5$ and $\Delta M=6$ vortex states for 6 electron quantum dot.}
\begin{tabular}{l l l l l}
&$M$ & $|\Phi\rangle        $   & $\langle\Psi|\Phi\rangle$ & $\langle\Psi|\Phi_{\rm CM}\rangle$\\
\hline
$d=l$&20 & $|1011111\rangle$        &       0.83                &         0.97                     \\
&25 & $|10011111\rangle$       &       0.67                &         0.80                     \\
&30 & $|100011111\rangle$       &       0.61                &         0.79                     \\
&35 & $|1000011111\rangle$      &       0.57                &         0.77                     \\
&40 & $|10000011111\rangle$     &       0.53                &         0.74                     \\
&45 & $|100000011111\rangle$    &       0.38                &         0.57                     \\
\hline
$d=3l$&21 & $|0111111\rangle$         &       0.92                &         0.996                    \\
&27 & $|00111111\rangle$        &       0.84                &         0.97                     \\
&33 & $|000111111\rangle$       &       0.75                &         0.92                     \\
&39 & $|0000111111\rangle$      &       0.66                &         0.85                     \\
&45 & $|00000111111\rangle$     &       0.53                &         0.72                     \\
\hline
\end{tabular}
\end{table}

The ground states with $M=M_{\rm MDD}+kN$ can be approximated by the (unnormalized) trial wave function
$\Phi_{\rm CM}=\prod(z_i-z_j)\prod(z_i-z_{\rm CM})^k$. This is obtained by eliminating the center-of-mass motion by the transformation\cite{harju} $z_i\mapsto z_i-z_{\rm CM}$ from the $\nu=1$ integer Hall effect state with $k$ holes in the origin. The approximation scheme is generalized to arbitrary angular momenta $M$ by taking the highest weight many-body basis vector $|\Phi\rangle$ of the ground state and performing the transformation  $z_i\mapsto z_i-z_{\rm CM}$, $|\Phi\rangle\mapsto |\Phi_{\rm CM}\rangle$. We denote the basis state with quantum numbers $m_1,\ldots,m_N$ by a sequence, in which the elements number $m_1,\ldots,m_N$ are $1$ and the rest are zero. The thus obtained trial wave functions and their overlap with the ground states are presented in Table I. At $d=l$, the overlaps are rather high until $M=M(\nu=\frac{1}{3})=45$, which is still better described by the corresponding Laughlin state trial wave function (overlap $\sim0.97$). At $d=3l$, the overlaps are higher and even $M=M(\nu=\frac{1}{3})$ is slightly better described by our trial wave function than as a Laughlin state ($\sim0.67$). The two rightmost diagrams in Fig.~\ref{cwf} with $d=l$ (upper) and $d=3l$ (lower) both having $M(\nu=\frac{1}{3})$ also illustrate this break-down of the incompressible Laughlin state in thick systems as the vortices are gradually less bound to the electron co-ordinates. This effect due to the effective interaction is in contrast to the opposite effect caused by a screened Coulomb interaction\cite{stopa} whereby, in the strong screening limit, the zeros are exactly localized to the electron positions. Both cases are, of course, direct consequence of Haldane's pseudopotential parameters of the corresponding interactions and the fact that the Laughlin state is exactly obtained for an interaction having only the first pseudopotential coefficient nonzero. \cite{haldane}

The conclusion to be drawn from Table I is that for the $\Delta M=N$ the hole is created or extended at the origin, while for $\Delta M=N-1$ states the essential mechanism of increasing the angular momentum is to create a hole next to the origin. Similarly, for the configurations with two electrons at the center realized in the eight electron dot, the trial wave functions $|11000111111\rangle_{\rm CM}$ and $|110000111111\rangle_{\rm CM}$ may be applicable.

\section{Summary}
\label{sec:conclusions}
We have studied the phase diagram of spin-polarized quantum dots in high magnetic fields as a function of the thickness, and found roughly that for electron number $N$ equal to six and seven, the ground state angular momenta at different thickness occur at either intervals $N$ or $N-1$. For $N\leqslant$ 5, only the intervals $N$ occur, while for $N\geqslant8$ other intervals of type $N-k$ can occur. Moreover, we have interpreted the angular momenta in terms of vortex formation near the center of the quantum dot and constructed trial wave functions that have a moderate overlap with the exact solutions.
\section{Appendix}

The formula for the interaction matrix elements in the effective oscillator units obtained by transforming into relative and center-of-mass co-ordinates reads

\begin{equation}
\begin{split}
&M^l_{mn}=\frac{l_*}{a_*}[(m+l)!n!m!(n+l)!]^{-\frac{1}{2}}\\
&\times\sum_{\eta=0}^{m+l}\sum_{\zeta=0}^{m}\sum_{\xi=0}^{n+l}
\theta(n+l-\eta-\zeta+\xi)\theta(\eta+\zeta-\xi)(-1)^{\eta-\xi}\\
&\times\tbinom{m+l}{\zeta}\tbinom{n}{\eta}\tbinom{m}{\xi}\tbinom{n+l}{\zeta+\eta-\xi}
(m+n+l-\eta-\zeta)!\\
&\times[\tfrac{1}{\sqrt{2}}\Gamma(\eta+\zeta+\tfrac{1}{2}) _1F_1(\tfrac{1}{2},\tfrac{1}{2}-\eta-\zeta,\tfrac{d^2}{2})+\tfrac{(\eta+\zeta)!}{\sqrt{\pi}2^{\eta+\zeta+1}}\\
&\times\Gamma(-\eta-\zeta-\tfrac{1}{2}) _1F_1(\eta+\zeta+1,\eta+\zeta+\tfrac{3}{2},\tfrac{d^2}{2})]\ ,\\
\end{split}
\end{equation}
where $\theta$ is the step function, $_1F_1$ is the Kummer confluent hypergeometric function, and $l_*$ and $a_*$ are the effective oscillator length and the effective Bohr radius $a_*=\frac{4\pi\epsilon\hbar^2}{m^*e^2}$.

\end{document}